\documentclass[conference]{IEEEtran}
\IEEEoverridecommandlockouts
\usepackage{cite}
\usepackage{amsmath,amssymb,amsfonts}
\usepackage{algorithmic}
\usepackage{graphicx}
\usepackage{textcomp}
\usepackage{xcolor}
\usepackage{multirow}
\usepackage{setspace}
\usepackage{microtype}
\usepackage{letterspace}
\usepackage[utf8]{inputenc}
\usepackage{subcaption}
\usepackage{caption}
\def\BibTeX{{\rm B\kern-.05em{\sc i\kern-.025em b}\kern-.08em
    T\kern-.1667em\lower.7ex\hbox{E}\kern-.125emX}}
\begin{document}
\begin{spacing}{0.97}
\title{
{\fontsize{23.8pt}{10pt}\selectfont A 28.6 mJ/iter Stable Diffusion Processor for Text-to- Image Generation with Patch Similarity-based Sparsity Augmentation and Text-based Mixed-Precision*} \\
%A 28.6 mJ/iter Stable Diffusion Processor for Text-to- Image Generation with Patch Similarity- based Sparsity Augmentation and Text-based Mixed-Precision \\
\thanks{*2024 IEEE International Symposium on Circuits and Systems (ISCAS)}
}

\author{\large Jiwon Choi, Wooyoung Jo, Seongyon Hong, Beomseok Kwon, Wonhoon Park, and Hoi-Jun Yoo\\
KAIST, Daejeon, Republic of Korea (E-mail: jw.choi@kaist.ac.kr)}
\end{spacing}
\IEEEaftertitletext{\vspace{-0.5\baselineskip}}
\maketitle

\begin{abstract}
This paper presents an energy-efficient stable diffusion processor for text-to-image generation. While stable diffusion attained attention for high-quality image synthesis results, its inherent characteristics hinder its deployment on mobile platforms. The proposed processor achieves high throughput and energy efficiency with three key features as solutions: 1) Patch similarity-based sparsity augmentation (PSSA) to reduce external memory access (EMA) energy of self-attention score by 60.3 \%, leading to 37.8 \% total EMA energy reduction. 2) Text-based important pixel spotting (TIPS) to allow 44.8 \% of the FFN layer workload to be processed with low-precision activation. 3) Dual-mode bit-slice core (DBSC) architecture to enhance energy efficiency in FFN layers by 43.0 \%. The proposed processor is implemented in 28 nm CMOS technology and achieves 3.84 TOPS peak throughput with 225.6 mW average power consumption. In sum, 28.6 mJ/iteration highly energy-efficient text-to-image generation processor can be achieved at MS-COCO dataset.\\
\end{abstract}


\section{Introduction}
Demands for generative AI have surged in recent years due to its outstanding quality and potential for creativity. Extensive interest has been drawn to image synthesis, where alternatives to traditional generative models have been proposed, resulting in high-quality results and better generalizability \cite{b1,b2,b3,b4,b5}. Stable Diffusion (SD) has emerged as a powerful solution in image synthesis and is widely adopted for commercial use \cite{b6}. SD excels in generating images based on text descriptions, enabling users to obtain their intended pictures from text queries. 

\begin{figure}[t]
    \begin{subfigure}[b]{0.5\textwidth}
        \centerline{\includegraphics[width=0.98\columnwidth]{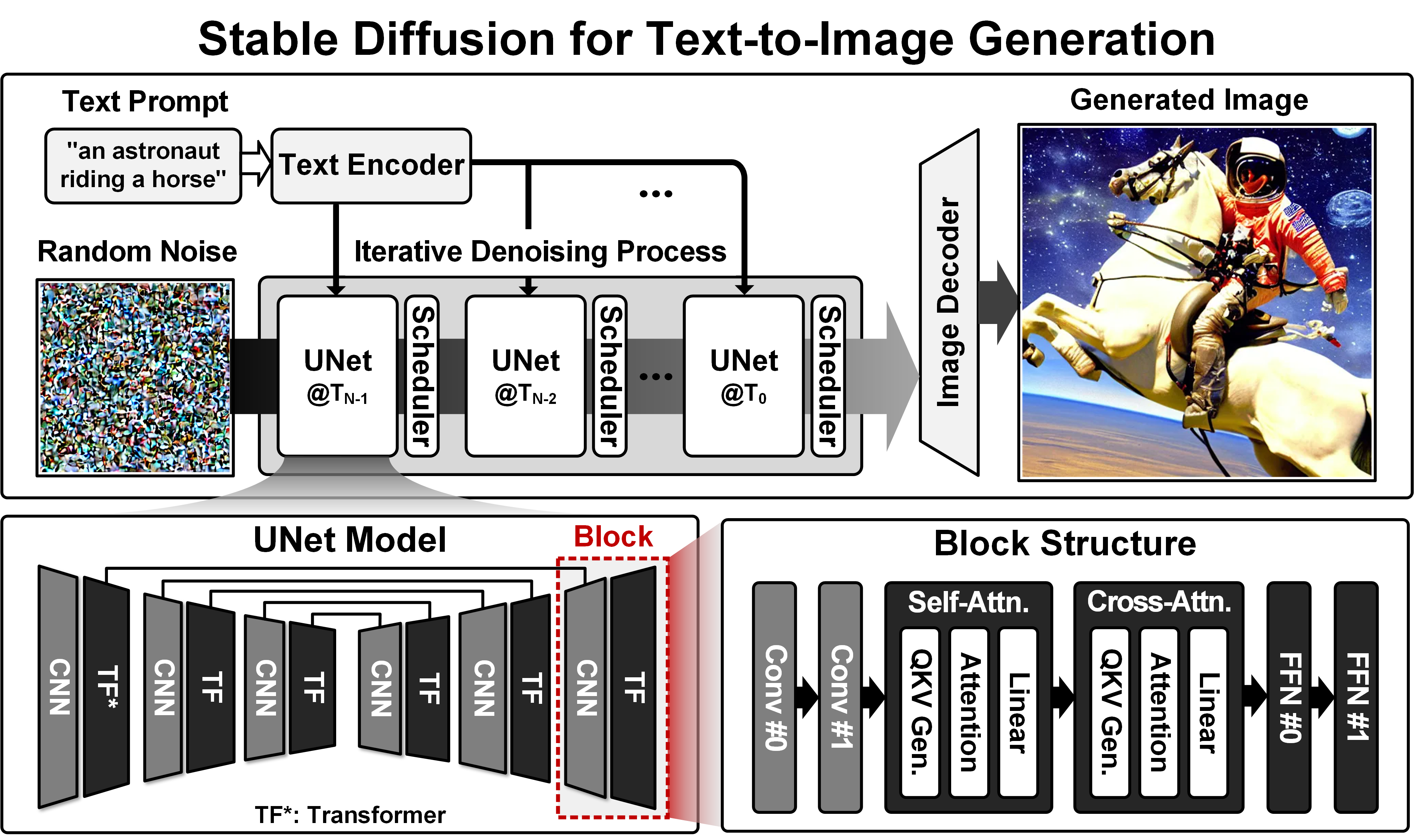}}
        \vspace{-1mm}
        \caption{}
        \label{fig1-1}
    \end{subfigure}
    \hfill
    \begin{subfigure}[b]{0.5\textwidth}
        \centerline{\includegraphics[width=0.98\columnwidth]{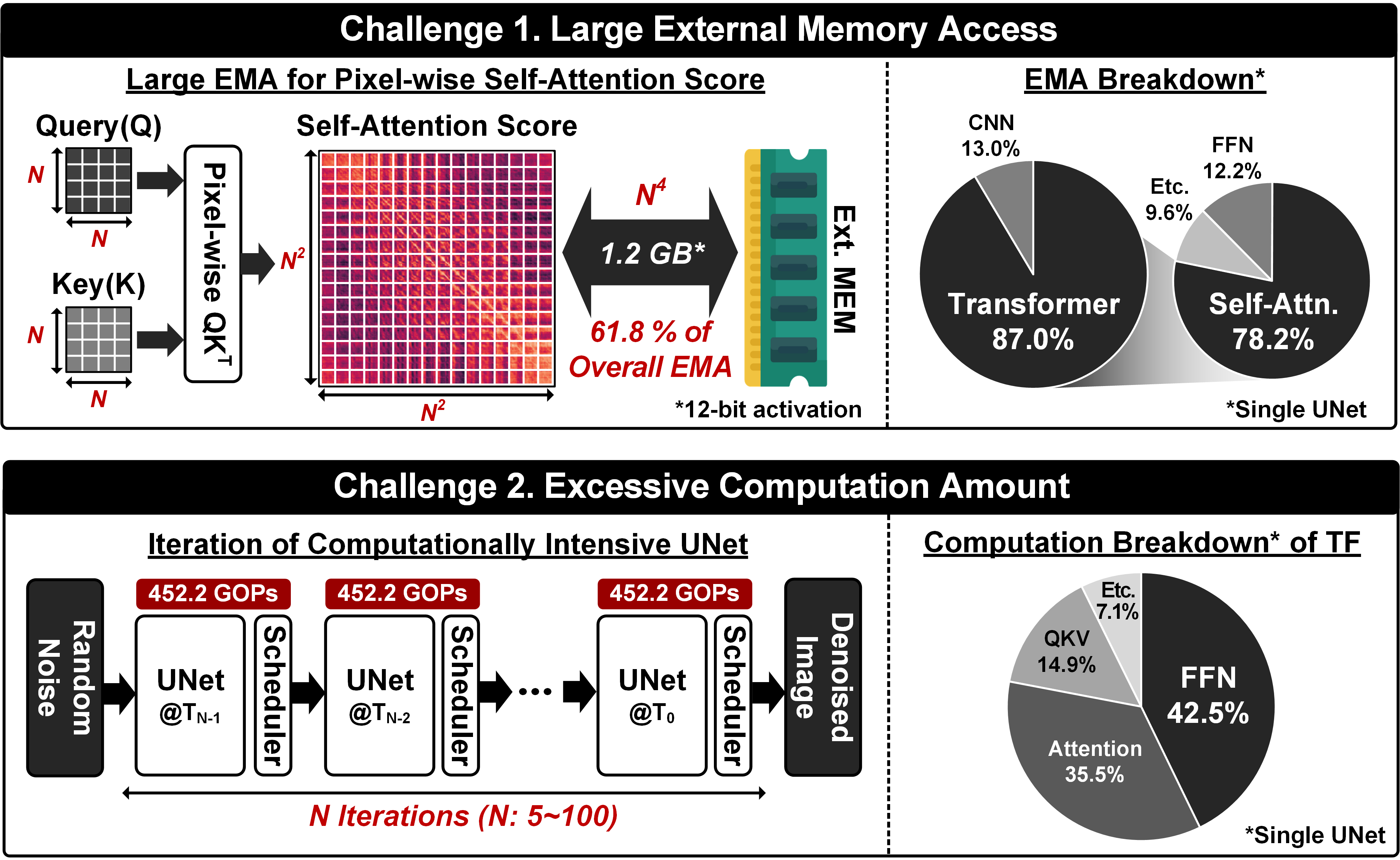}}
        \vspace{-1mm}
        \caption{}
        \label{fig1-2}
    \end{subfigure}
    \vspace{-6mm}
    \caption{(a) Overview of stable diffusion model. (b) Two main challenges.}
    \label{fig1}
\vspace{-7mm}
\end{figure}

Fig.~\ref{fig1}(a) shows the overall processing flow of SD for text-to-image generation. SD comprises three main stages: 1) text encoding, 2) iterative denoising process for image generation, and 3) image decoding to acquire the final image.  The UNet employed in SD consists of multiple blocks and residual connections. Each block comprises a convolutional neural network (CNN) and transformer stages. CNN stage involves two CNN layers, whereas the transformer stage contains self-attention, cross-attention, and feedforward network (FFN) layers. Cross-attention incorporates text information in the denoising process  by means of attention computation.

Although utilizing UNet as a backbone demonstrates remarkable image quality by alternating CNN and transformer stages, its inherent model architecture characteristics hinder mobile devices from deployment. To efficiently employ SD at mobile devices, two main challenges should be addressed in terms of memory and computation requirements. First, attention computation at the transformer stage incurs significant amounts of external memory access (EMA). Even a single iteration of UNet requires 1.9 GB of EMA with 12-bit input activation and 8-bit weight precision. 87.0 \% of the total EMA is required for the transformer stage in a single UNet, where self-attention layers occupy 78.2 \% of the EMA in the transformer. Since self-attention is performed pixel-wise to capture the spatial correlation in an image, its self-attention score necessitates an exponentially explosive EMA, which accounts for 61.8 \% of the overall EMA (Fig.~\ref{fig1}(b)). Secondly, excessive computation amount is required due to the iterations of computationally intensive UNet. While CNN and transformer divide the overall computational workload in a similar proportion, optimization with mixed-precision schemes for CNN has been widely investigated by the prior arts \cite{b7}. However, transformer block in UNet, especially for FFN layers following the cross-attention layers, still remains a challenge. Fig.~\ref{fig1}(b) shows the challenge posed by the excessive computation amount of SD and the dominant computation amount of an FFN layer, which accounts for 42.5\% of the transformer stages. Thus, it is crucial to reduce the computation workload of the FFN layer. 

\begin{figure}[tbp]
\captionsetup{justification=raggedright,singlelinecheck=false} % 여기에 설정을 적용
\centerline{\includegraphics[width=0.98\columnwidth]{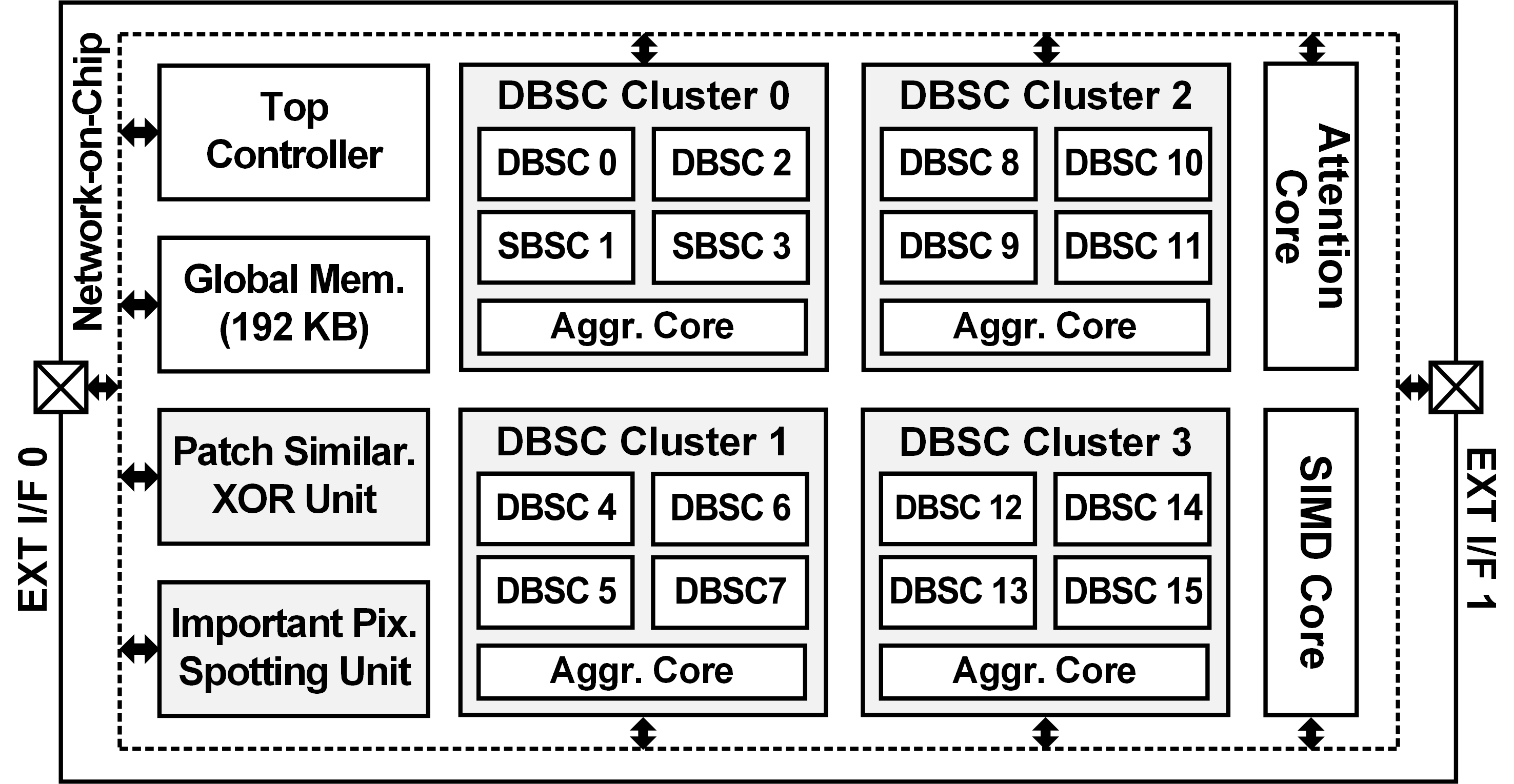}}
\caption{Overall architecture.}
\label{fig2}
\vspace{-7mm}
\end{figure}

To the best of our knowledge, there are no existing approaches to designing a SD processor for mobile devices. We present an energy-efficient SD processor with three key features: 
1) Patch similarity-based sparsity augmentation (PSSA) to reduce the EMA energy of self-attention score by 60.3 \%. 
2) Text-based important pixel spotting (TIPS) for allocating 44.8 \% of FFN layers workload to low-precision. 
3) Dual-mode bit-slice core (DBSC) for mixed-precision computation and dual stationary modes to support both CNN and transformer layers, which increases energy efficiency in the FFN layer by 43.0 \%. As a result, the proposed SD processor was verified in a 28 nm CMOS process and achieved energy efficiency of 28.6 mJ/iteration with only 0.002 CLIP loss and 0.16 FID loss on the MS-COCO dataset.

\section{Overall Architecture}
Fig.~\ref{fig2} shows the overall architecture of the proposed SD processor. It consists of 4 dual-mode bit-slice core (DBSC) clusters, a patch similarity-based XOR unit (PSXU), an important pixel spotting unit (IPSU), a 192 KB global memory, a top controller, an attention core, a SIMD core, and 2-D mesh-type network-on-chip (NoC). The DBSC cluster comprises 4 DBSCs and an aggregation core. Each DBSC includes a 16×16 processing element (PE) array, along with 6 KB of input memory, 2.25 KB of weight memory, and 12 KB of output memory. The DBSC supports a dual stationary mode to optimize reusability. It supports input stationary for the CNN stage and weight stationary for the transformer stage. The aggregation core accumulates partial sums generated from the 4 DBSCs to generate the final output. The PSXU effectively compresses self-attention scores for EMA reduction. The IPSU identifies important pixels based on the given text, enabling mixed-precision FFN. The attention core supports input skipping with a CSR decoder for processing the attention layer efficiently. The SIMD core performs activation functions, on-chip quantization, and group normalization.

\begin{figure}[tp]
    \begin{subfigure}[b]{\columnwidth}
        \centerline{\includegraphics[width=0.98\textwidth]{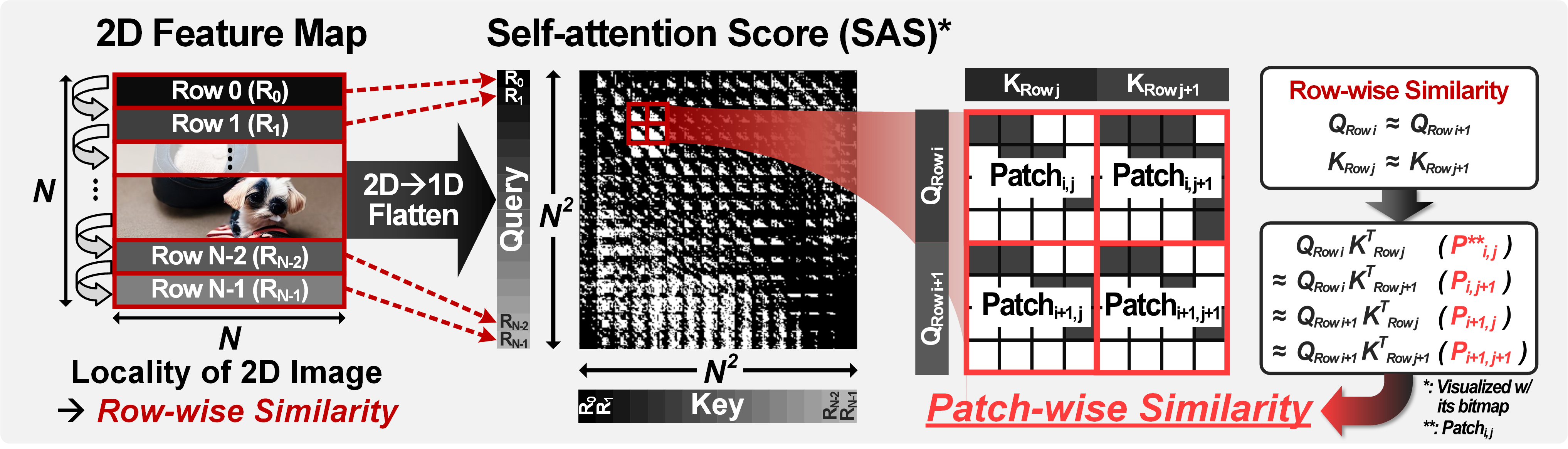}}
        \vspace{-1mm}
        \caption{}
        \label{fig3-1}
    \end{subfigure}
    \hfill
    \begin{subfigure}[b]{\columnwidth}
        \centerline{\includegraphics[width=0.98\textwidth]{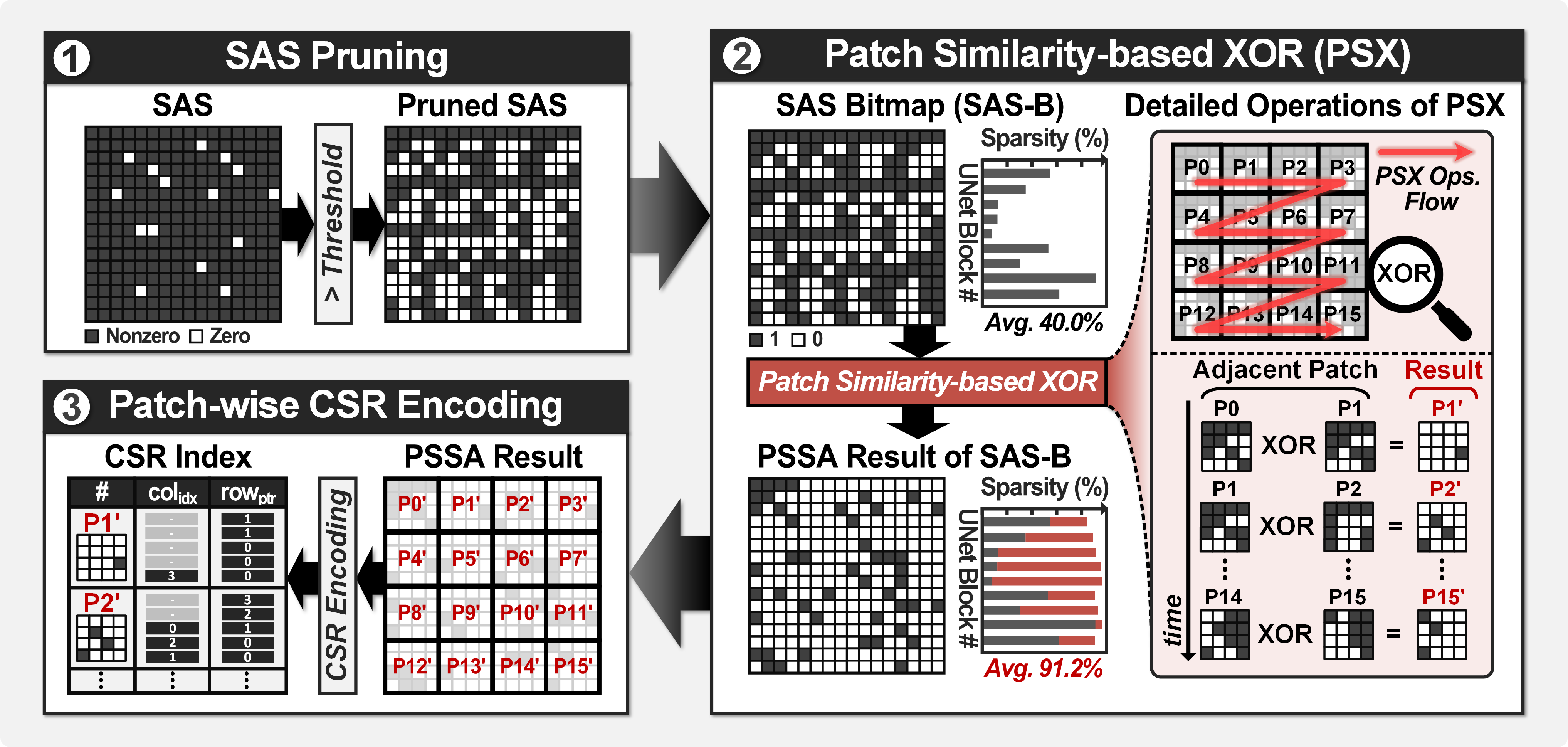}}
        \vspace{-1mm}
        \caption{}
        \label{fig3-2}
    \end{subfigure}
    \caption{(a) Patch-wise similarity in SAS. (b) Compression flow of SAS with proposed patch similarity-based sparsity augmentation.}
    \label{fig3}
    \vspace{-7mm}
\end{figure}

\section{Effective Compression of Self-attention Score}

\subsection{Self-attention Score Bitmap Sparsity Augmentation}

The self-attention mechanism in SD operates with pixel-wise embeddings, resulting in a significant size of the self-attention scores (SAS). Reducing the SAS EMA is imperative to enable SD processing on mobile devices since SAS occupies 61.8 \% of the total EMA. Fortunately, pixel-wise self-attention provides an opportunity for efficient compression by utilizing local similarities within SAS. Fig.~\ref{fig3}(a) illustrates the patch-wise similarity in SAS. Owing to the spatial locality among pixels, the 2D input feature map of a self-attention layer exhibits similarities between adjacent rows. When the input feature map is transformed into a query and a key, it is flattened row-wise into a 1D format for pixel-wise self-attention. Consequently, SAS patches created through the attention between a row of the query and a row of the key exhibit similarities between adjacent patches. 

Fig.~\ref{fig3}(b) shows the overall compression flow of SAS with proposed patch similarity-based sparsity augmentation (PSSA). The compression process consists of three steps. The first step is SAS pruning, which prunes SAS values using a predefined fixed threshold in an unstructured manner. The second step is patch similarity-based XOR. Since the pruned SAS and its bitmap exhibit patch-wise similarity, the XOR operation is applied to adjacent bitmap patches in the horizontal direction of the pruned SAS bitmap to increase the bitmap sparsity. The third step involves patch-wise compression using compressed sparse row (CSR) encoding. Experimental results demonstrated local CSR encoding for each patch yielded a higher compression rate compared to global CSR encoding for the entire SAS bitmap since the encoding overhead of CSR decreases with the target size.

\begin{figure}[tp]
\vspace{-5mm}
\captionsetup{justification=raggedright,singlelinecheck=false} % 여기에 설정을 적용
\centerline{\includegraphics[width=0.98\columnwidth]{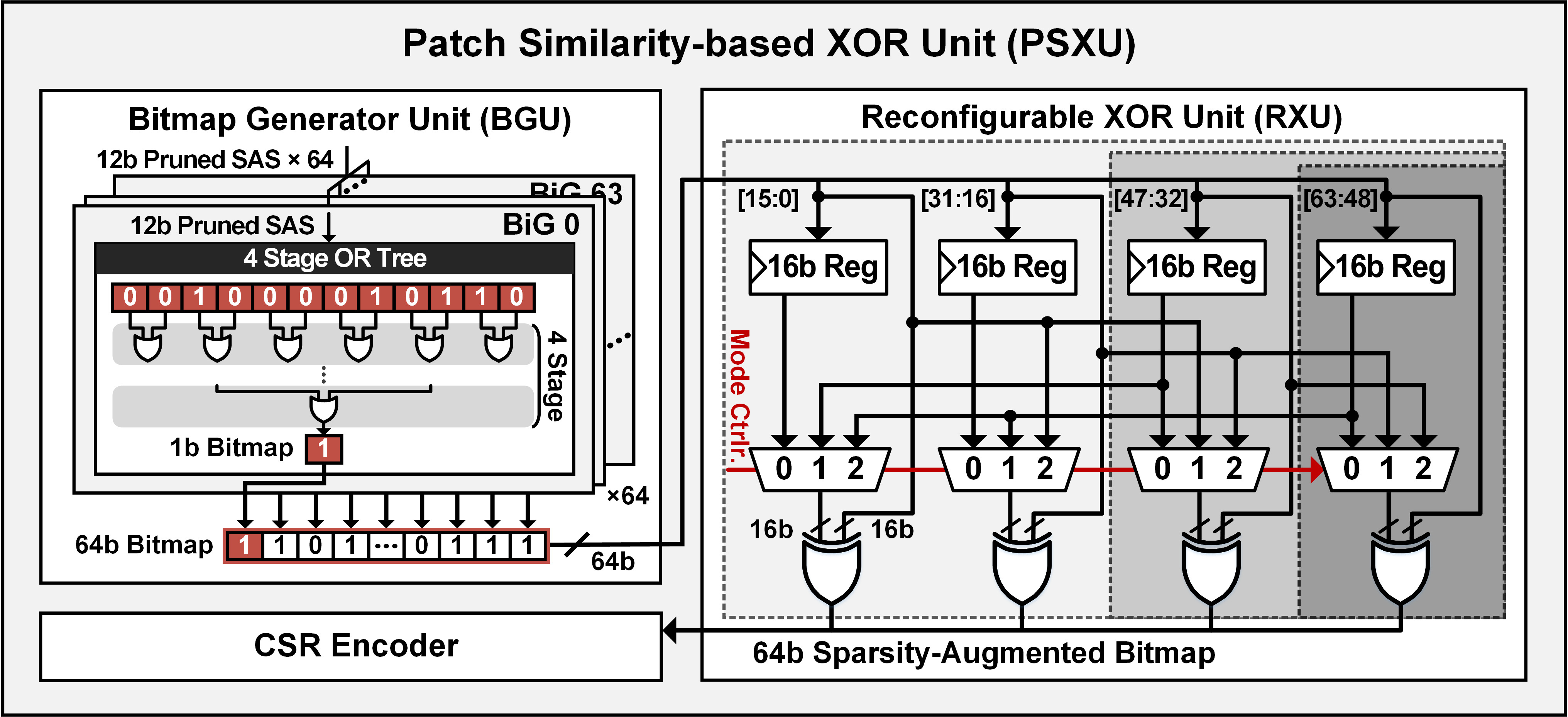}}
\caption{Proposed patch similarity-based XOR unit (PSXU).}
\label{fig4}
\vspace{-3mm}
\end{figure}

\begin{figure}[tp]
    \begin{subfigure}[b]{0.52\columnwidth}
        \vspace{-1mm}
        \centerline{\includegraphics[width=\textwidth]{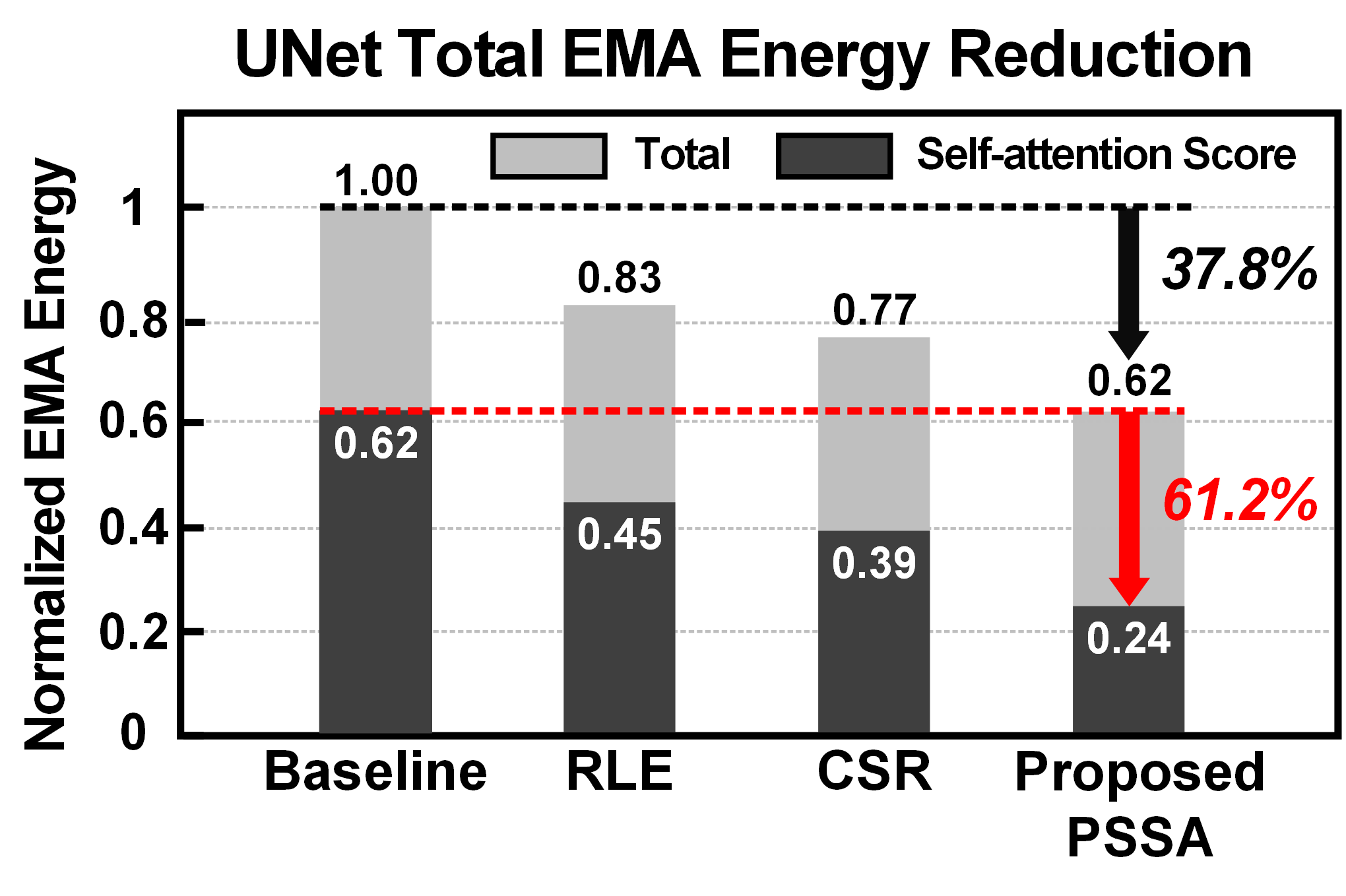}}
        \vspace{-2mm}
        \caption{}
        \label{fig5-1}
    \end{subfigure}
    \hspace{0mm}
    \begin{subfigure}[b]{0.43\columnwidth}
        \vspace{-1mm}
        \centerline{\includegraphics[width=\textwidth]{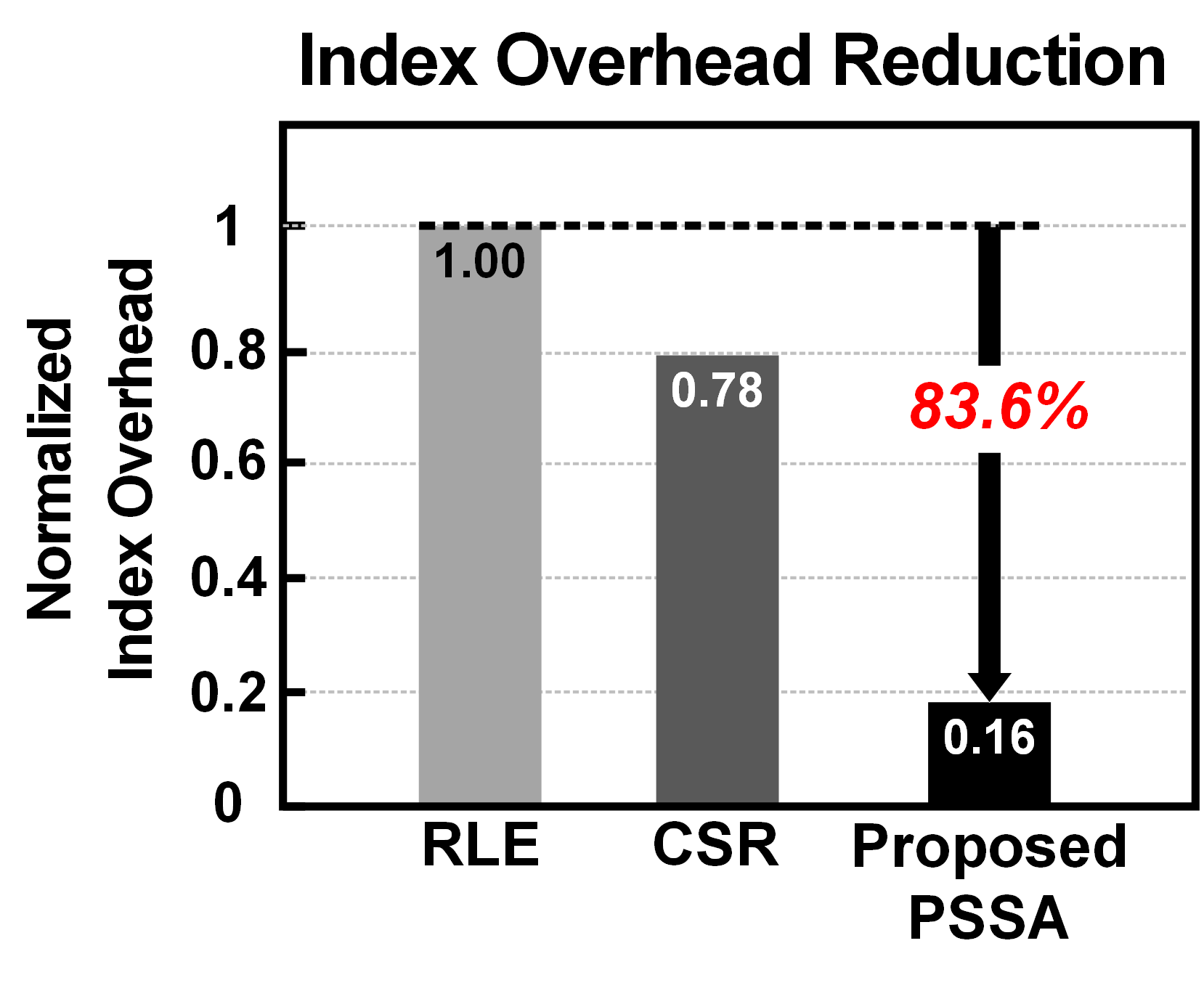}}
        \vspace{-2mm}
        \caption{}
        \label{fig5-2}
    \end{subfigure}
    \captionsetup{justification=raggedright,singlelinecheck=false} % 여기에 설정을 적용
    \vspace{-1mm}
    \caption{Performance of proposed PSSA.}
    \label{fig5}
    \vspace{-7mm}
\end{figure}

\subsection{Patch Similarity-based XOR Unit (PSXU)}
Fig.~\ref{fig4} shows the patch similarity-based XOR unit (PSXU) which enables the proposed PSSA. The PSXU comprises a bitmap generator unit (BGU) with 64 bitmap generators (BiGs), a reconfigurable XOR unit (RXU) to support different patch sizes, and a CSR encoder. Since each block in the UNet structure requires different token lengths for the query and key in the self-attention layer, the patch size varies with three possible cases: 16×16, 32×32, and 64×64. PSXU takes 64 data from one row of the SAS as an input, and BGU generates a 64-bit bitmap. The generated 64-bit bitmap is then transferred to an RXU. The mode control signal, which controls the four 3×1 MUXs in RXU, enables the RXU to operate reconfigurable as an XOR unit for a single 64-bit patch, two adjacent 32-bit patches, and four adjacent 16-bit patches. The sparsity-augmented bitmap generated in the RXU is fed into the CSR encoder to generate indices ($col_{idx}$, $row_{ptr}$).

Fig.~\ref{fig5} shows the performance analysis of the proposed PSSA with PSXU. The proposed PSSA reduced the EMA energy of SAS by 61.2 \%, 46.7 \%, and 38.5 \% compared to the baseline without any compression, conventional run-length encoding (RLE) encoding, and CSR encoding, respectively (Fig.~\ref{fig5}(a)). The total EMA energy of a single UNet was reduced by 37.8 \% compared to the baseline. In addition, the index overhead resulting from compression was reduced by 83.6 \% and 79.5 \% compared to RLE and CSR encoding, respectively as shown in Fig.~\ref{fig5}(b). 

\begin{figure}[t]
    \vspace{-5mm}
    \begin{subfigure}[b]{0.98\columnwidth}
        \centerline{\includegraphics[width=\textwidth]{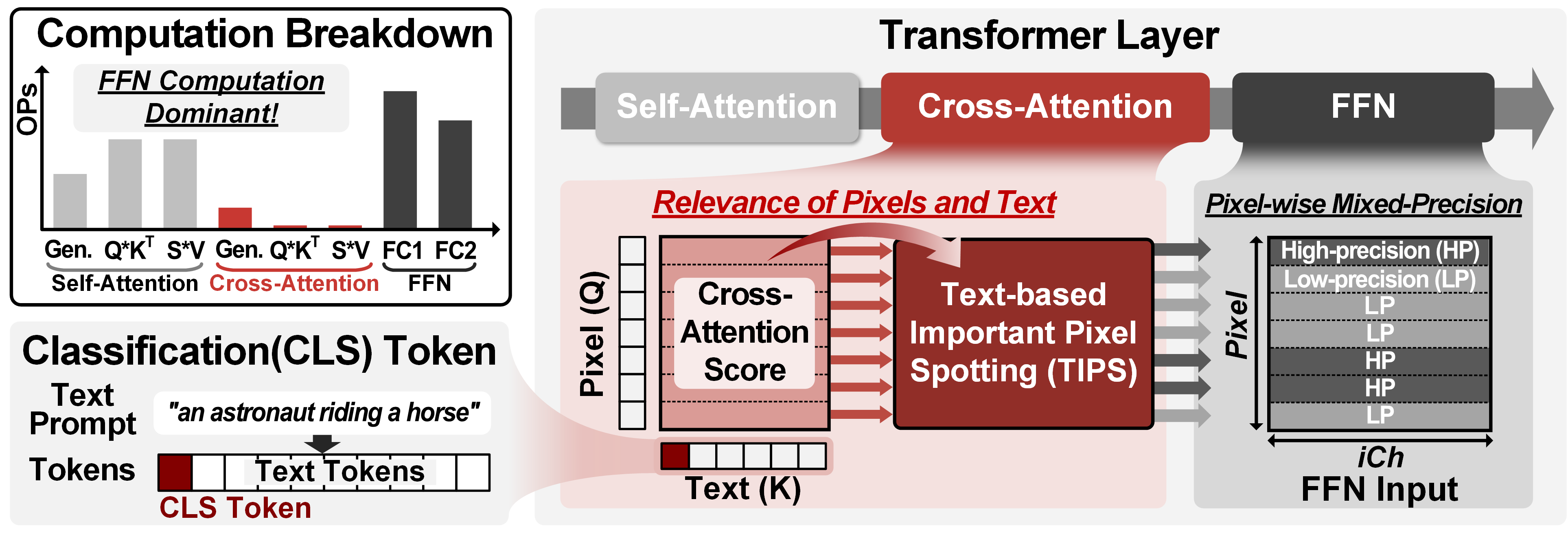}}
        \vspace{-1mm}
        \caption{}
        \label{fig6-1}
    \end{subfigure}
    \hfill
    \begin{subfigure}[b]{0.98\columnwidth}
        \centerline{\includegraphics[width=\textwidth]{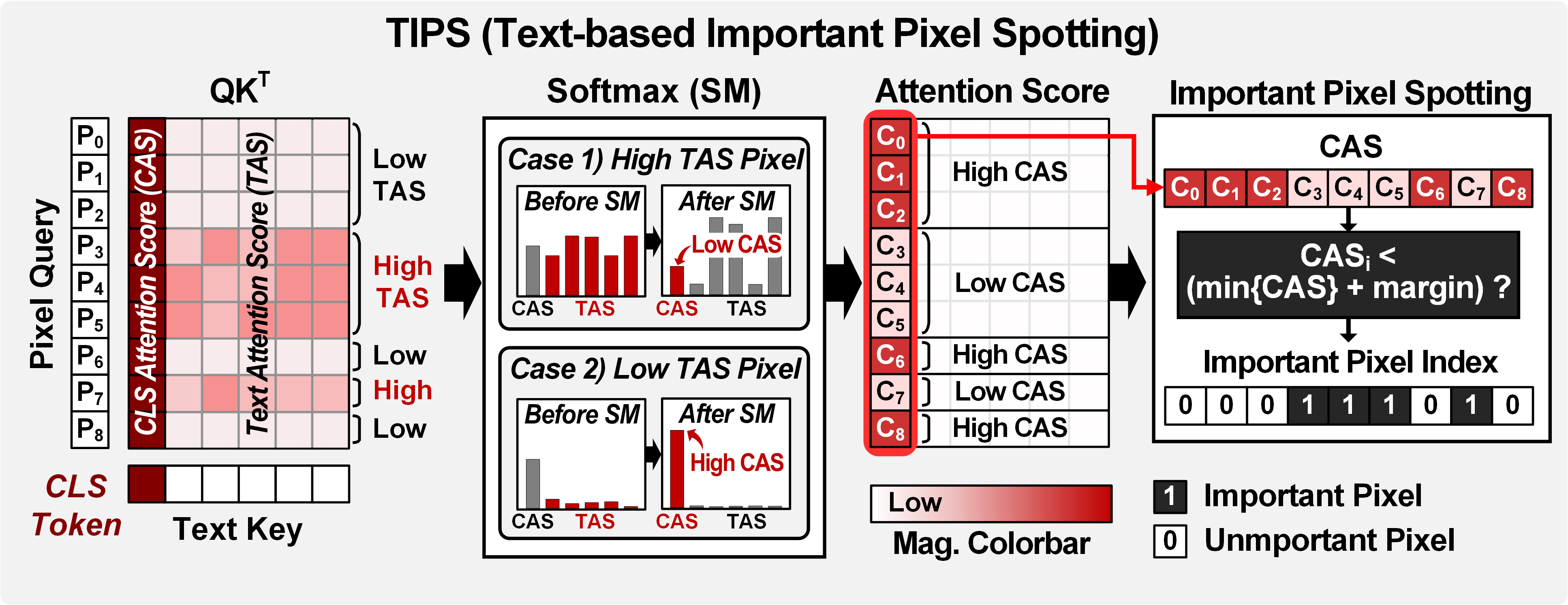}}
        \vspace{-1mm}
        \caption{}
        \label{fig6-2}
    \end{subfigure}
    \vspace{-1mm}
    \caption{(a) Pixel-wise mixed-precision in FFN using cross-attention score. (b) Proposed text-based important pixel spotting (TIPS).}
    \label{fig6}
    \vspace{-7mm}
\end{figure}

\section{Text-based Mixed-precision Processing}

\subsection{Text-based Important Pixel Spotting}
To reduce the dominating computation power of the FFN layer (Fig.~\ref{fig6}(a)), we focused on the cross-attention layer preceding the FFN layer. The proposed text-based important pixel spotting (TIPS) stems from the nature of cross-attention which decides the importance of pixels related to the given text. Pixels with have less importance can be allocated a lower precision after the cross-attention computation. This enables the mixed precision computation at the following FFN layers. Since there are no mixing operations between different pixel tokens in cross-attention and FFN layers, low-precision workloads can be effectively allocated through all FFN layers.

Fig.~\ref{fig6}(b) shows the detailed flow of proposed text-based important pixel spotting (TIPS). Text keys in cross-attention can be classified as the CLS token and text token, where the CLS token is positioned as the first token to capture the global context of a sentence \cite{b8, b9}. The cross-attention results for each pixel query and text keys fall into the CLS attention score (CAS) and text attention score (TAS). Pixel queries highly relevant to text keys tend to have higher TAS to reflect the details of the given text. Since the following softmax layer (SM) normalizes the score for each query, TAS and CAS scales are inversely proportional to each other (i.e. smaller CAS means higher TAS). By comparing the CAS for each pixel query, we can determine the relative importance of pixels with reference to the text. If the CAS of a pixel query is smaller than the predefined threshold, that pixel is spotted as an important pixel and its index is stored. Cross-attention results of spotted pixels are allocated a high precision (INT12). Pixels not spotted are regarded as unimportant, and their cross-attention results are allocated with low precision (INT6). Thanks to TIPS, lower precision can be allocated for  FFN inputs, enabling the mixed-precision processing of FFN layers.

\begin{figure}[tp]
\vspace{-5mm}
\captionsetup{justification=raggedright,singlelinecheck=false} % 여기에 설정을 적용
\centerline{\includegraphics[width=0.98\columnwidth]{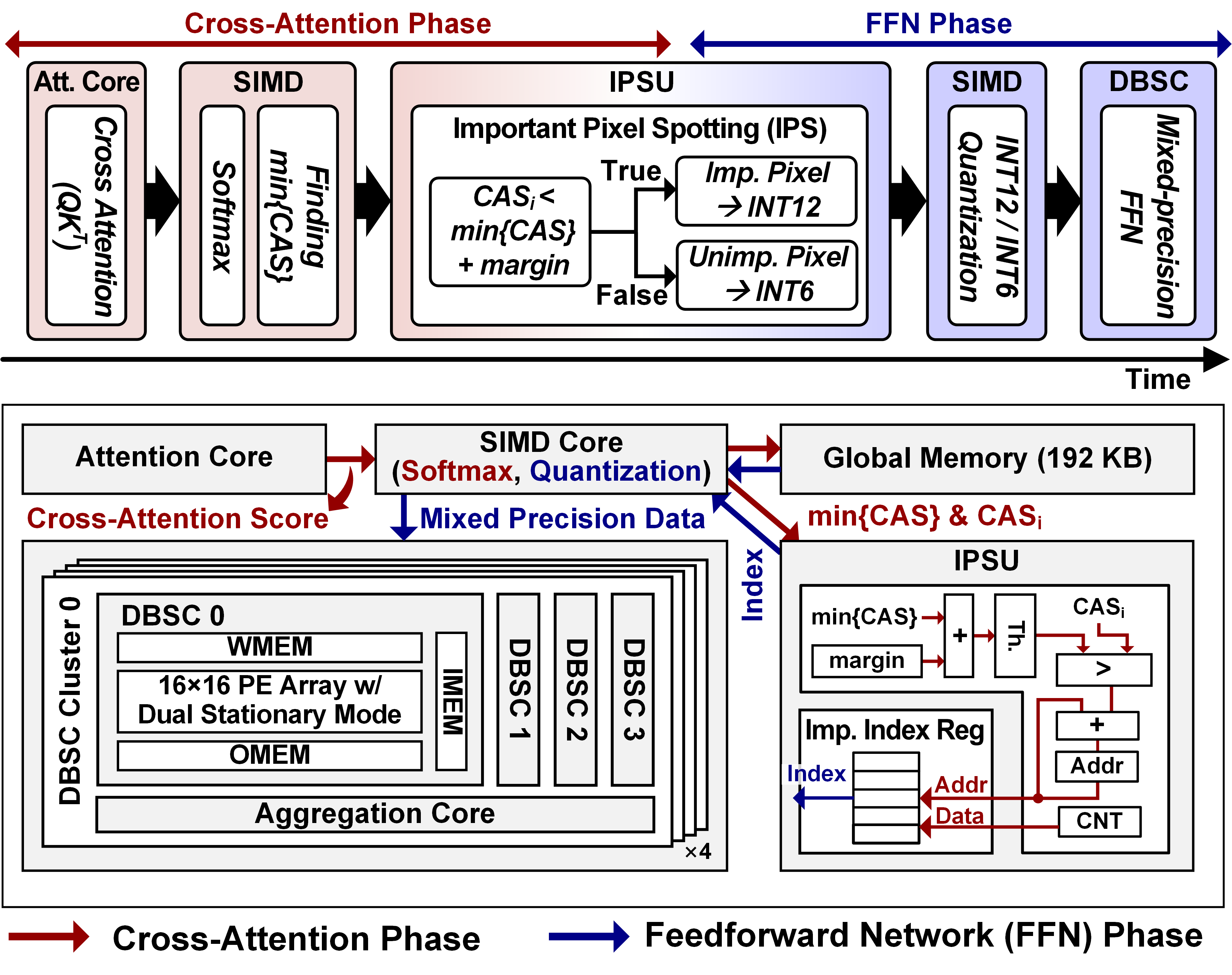}}
\vspace{-1mm}
\caption{Hardware Dataflow of TIPS.}
\label{fig7}
\vspace{-2mm}
\end{figure}

\begin{figure}[tp]
\vspace{-1mm}
\captionsetup{justification=raggedright,singlelinecheck=false} % 여기에 설정을 적용
\centerline{\includegraphics[width=0.98\columnwidth]{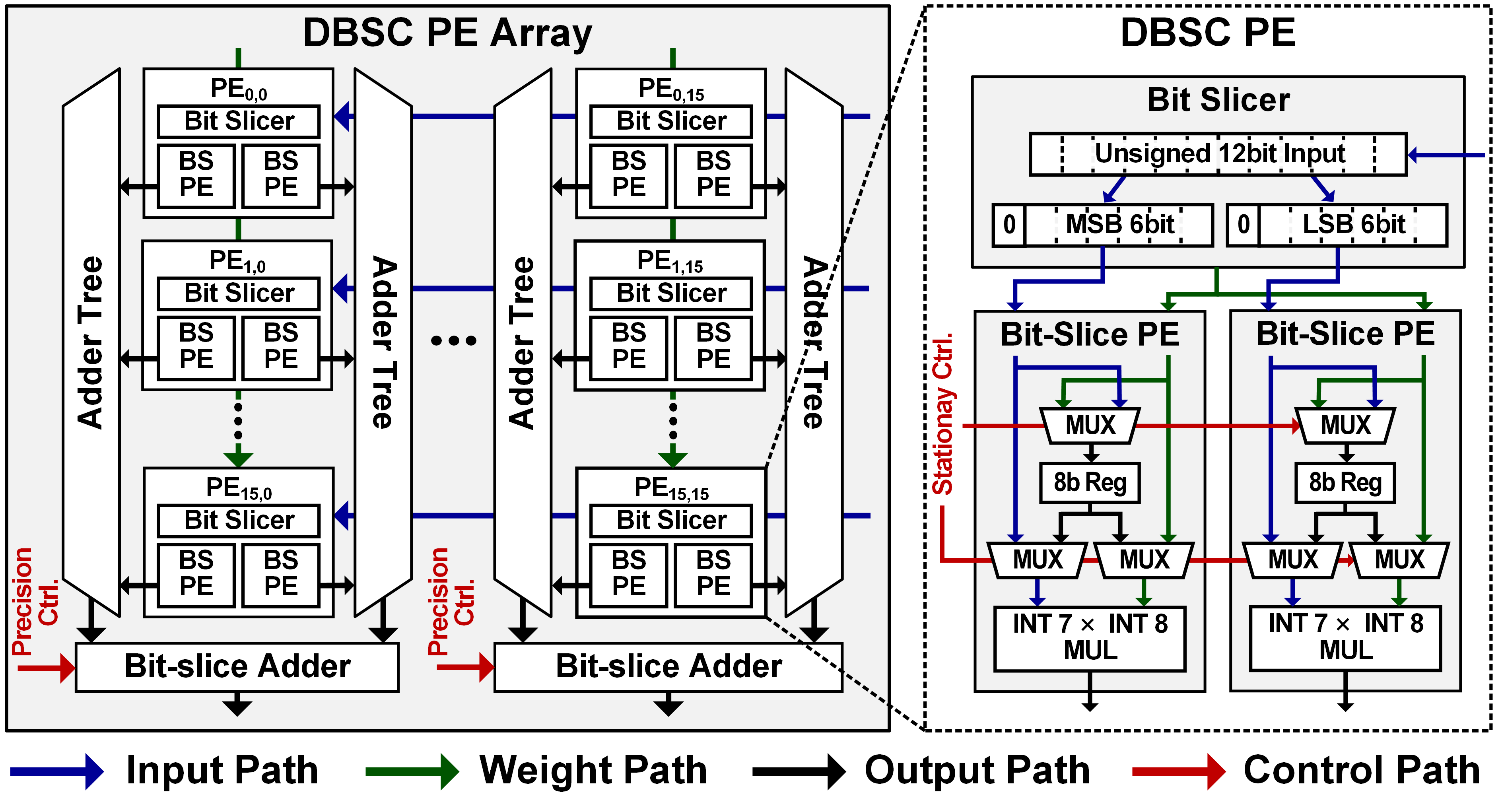}}
\vspace{-1mm}
\caption{PE array and PE architecture in DBSC.}
\label{fig8}
\vspace{-6mm}
\end{figure}

Fig.~\ref{fig7} illustrates the hardware dataflow across the cross-attention and FFN layer with TIPS. During the cross-attention phase, the cross-attention scores are calculated in the attention core and then transferred to the SIMD. In the SIMD core, SM operation followed by the minimum derivation of the CAS is performed. The cross-attention score is then transferred to the global memory, and CAS and minimum CAS are transferred to the IPSU. In IPSU, important pixels are spotted using the CAS and minimum CAS. During the FFN phase, the SIMD performs INT12/INT6 mixed-precision quantization using the indices of the important pixels received from the IPSU and the FFN activation from the global memory. The mixed-precision FFN activation is then transferred to the IMEM of each DBSC.

\begin{figure}[tp]
    \vspace{-5mm}
    \begin{subfigure}[b]{\columnwidth}
        \centering
        \includegraphics[width=\textwidth]{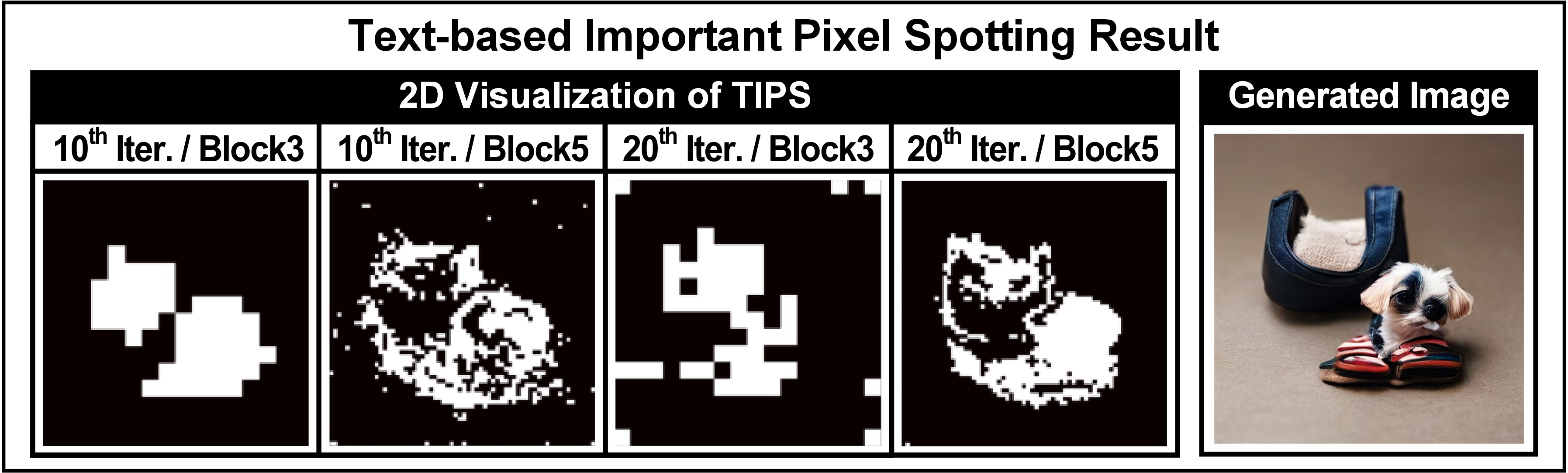}
        \caption{}
        \label{fig9-1}
    \end{subfigure}
    \hfill
    \begin{subfigure}[b]{0.595\columnwidth}
        \centering
        \includegraphics[width=\textwidth]{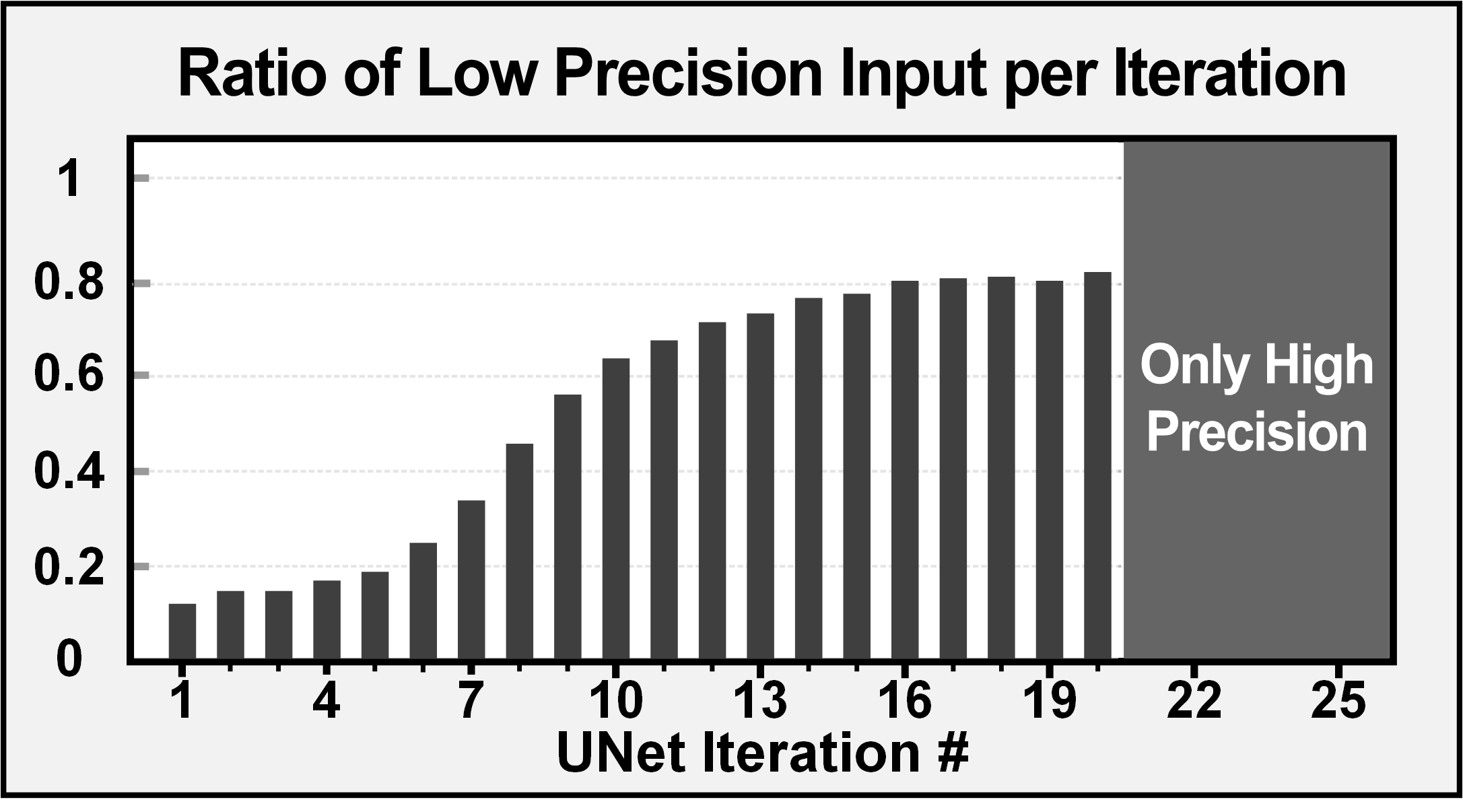}
        \caption{}
        \label{fig9-2}
    \end{subfigure}
    \hspace{0mm}
    \begin{subfigure}[b]{0.38\columnwidth}
        \centering
        \includegraphics[width=\textwidth]{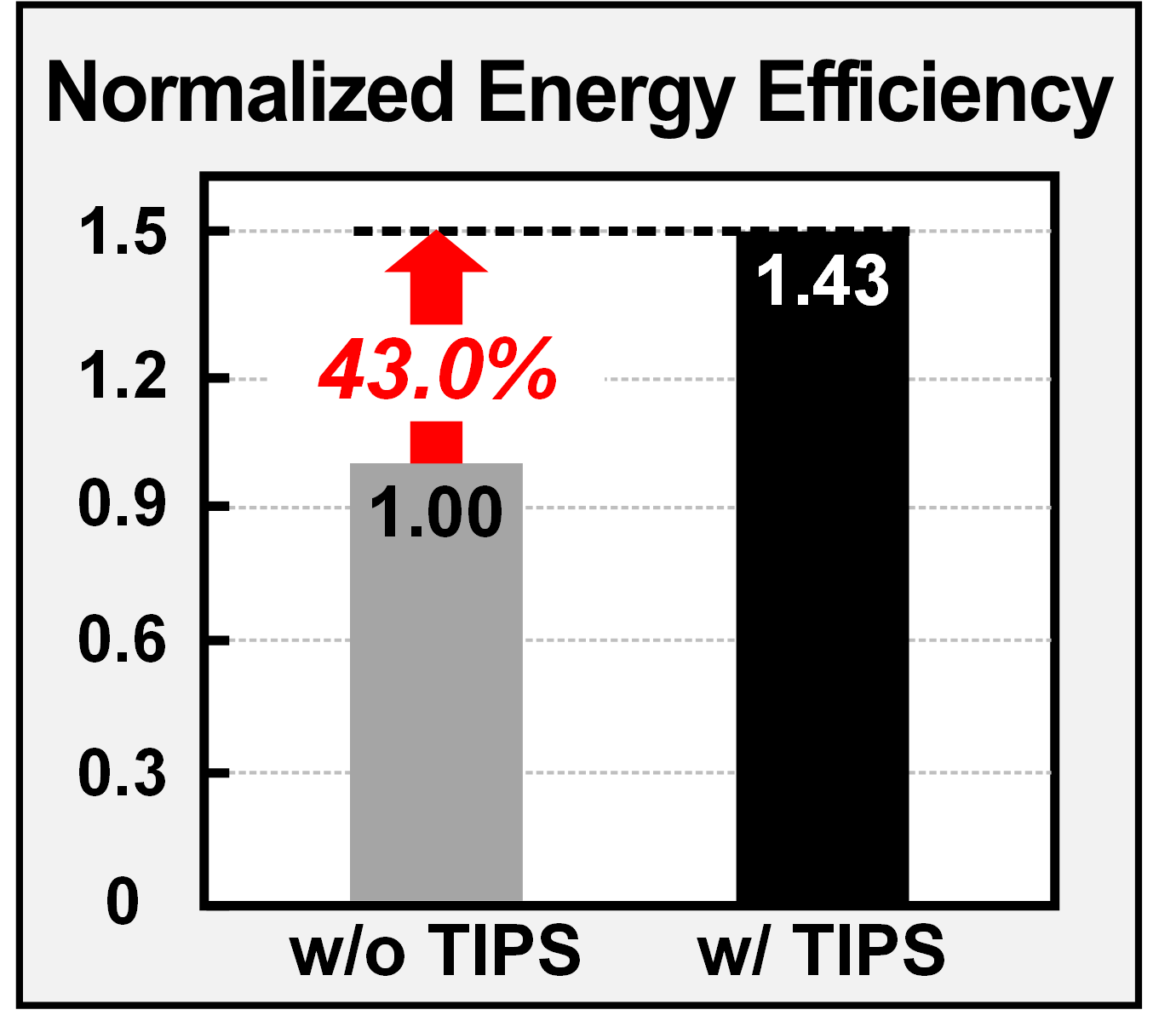}
        \caption{}
        \label{fig9-3}
    \end{subfigure}
    \caption{(a) 2D visualization of proposed TIPS results. (b) Low-precision input ratio for each iteration. (c) Energy efficiency of FFN layer with DBSC.}
    \label{fig9}
    \vspace{-3mm}
\end{figure}

The 2D visualizations of the spotted important pixels obtained by TIPS are compared with the final generated image in Fig.~\ref{fig9}(a). Note that the important pixels are spotted as 1 and appear white on the image, whereas the unimportant pixels are spotted as 0 and appear black. This comparison validates that the spotting results from TIPS effectively capture the relevance of each pixel to the given text. Fig.~\ref{fig9}(b) shows the ratio of low-precision data at each UNet iteration. The proposed TIPS was applied only to the first 20 out of 25 iterations due to quantization vulnerabilities observed in the last 5 iterations. 44.8 \% of pixels can be computed with low precision at the FFN layers, owing to TIPS.

\begin{figure}[tp]
\captionsetup{justification=raggedright,singlelinecheck=false} % 여기에 설정을 적용
\centerline{\includegraphics[width=\columnwidth]{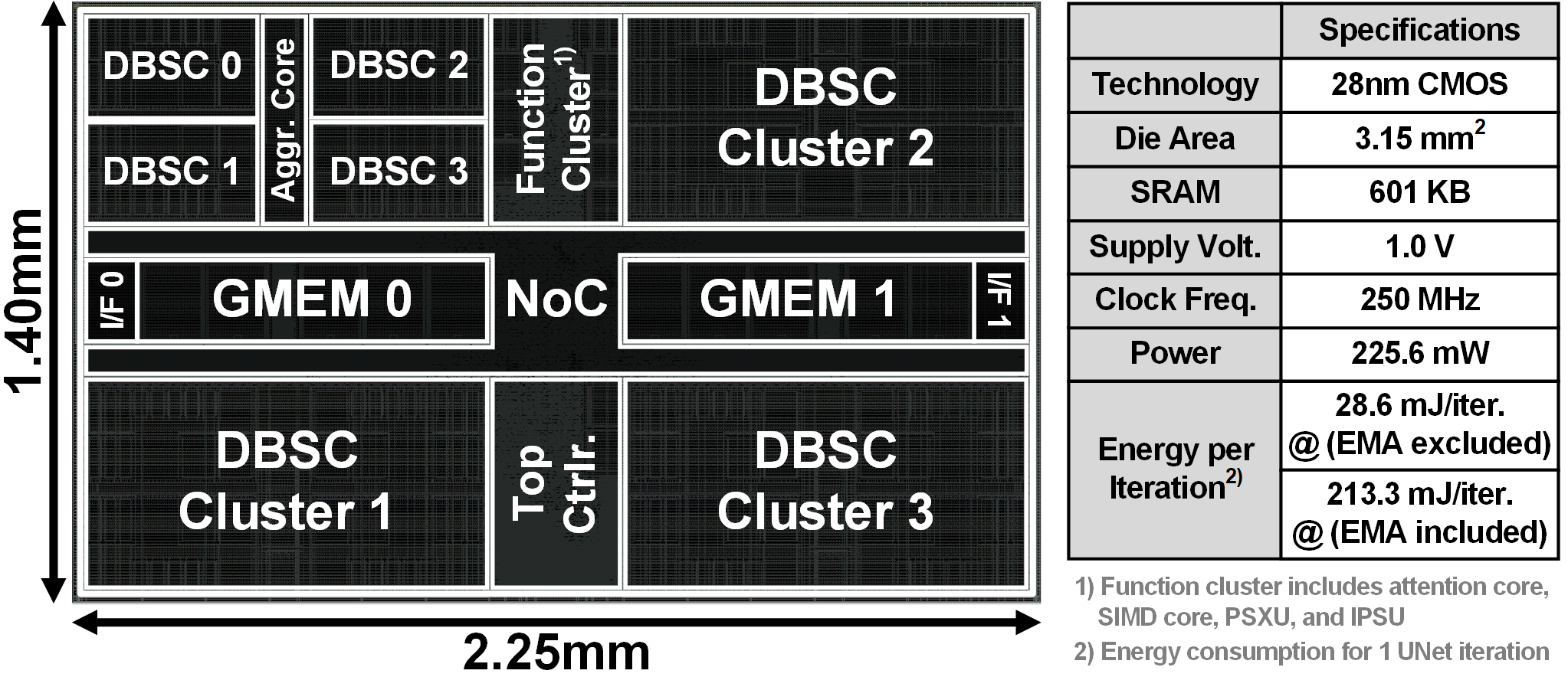}}
\vspace{-1mm}
\caption{Layout photograph and performance summary.}
\label{fig10}
\vspace{-6mm}
\end{figure}

\subsection{Dual-mode Bit-slice Core (DBSC) Architecture}
The proposed DBSC architecture supports mixed-precision computation along with input and weight stationary modes to optimize data reusability for both CNN and transformer layers (Fig.~\ref{fig8}). The PE array comprises 16 PE columns, and each PE column consists of 16 PEs, 2 adder trees, and a bit-slice adder. Each PE contains a bit slicer and two bit-slice PEs (BSPEs). The PE receives a 12-bit unsigned input and an 8-bit signed weight, and the input is then split into two 7-bit signed inputs by the bit slicer, with each being fed into its respective BSPE. The BSPE supports both input stationary and weight stationary modes and includes a multiplier for INT7 input and INT8 weight. Within a single PE column, the outputs from BSPEs in the same direction (left or right) are added together in a single adder tree. The outputs of the two adder trees in a PE column are added directly or after shifting, depending on whether the input is high precision or low precision. The proposed DBSC increases the energy efficiency in the FFN layer by 43.0 \% compared to the baseline that assumes INT12 for all inputs (Fig.~\ref{fig9}(c)).

\section{Implementation Results}
\begin{figure}[tp]
\captionsetup{justification=raggedright,singlelinecheck=false} % 여기에 설정을 적용
\centerline{\includegraphics[width=\columnwidth]{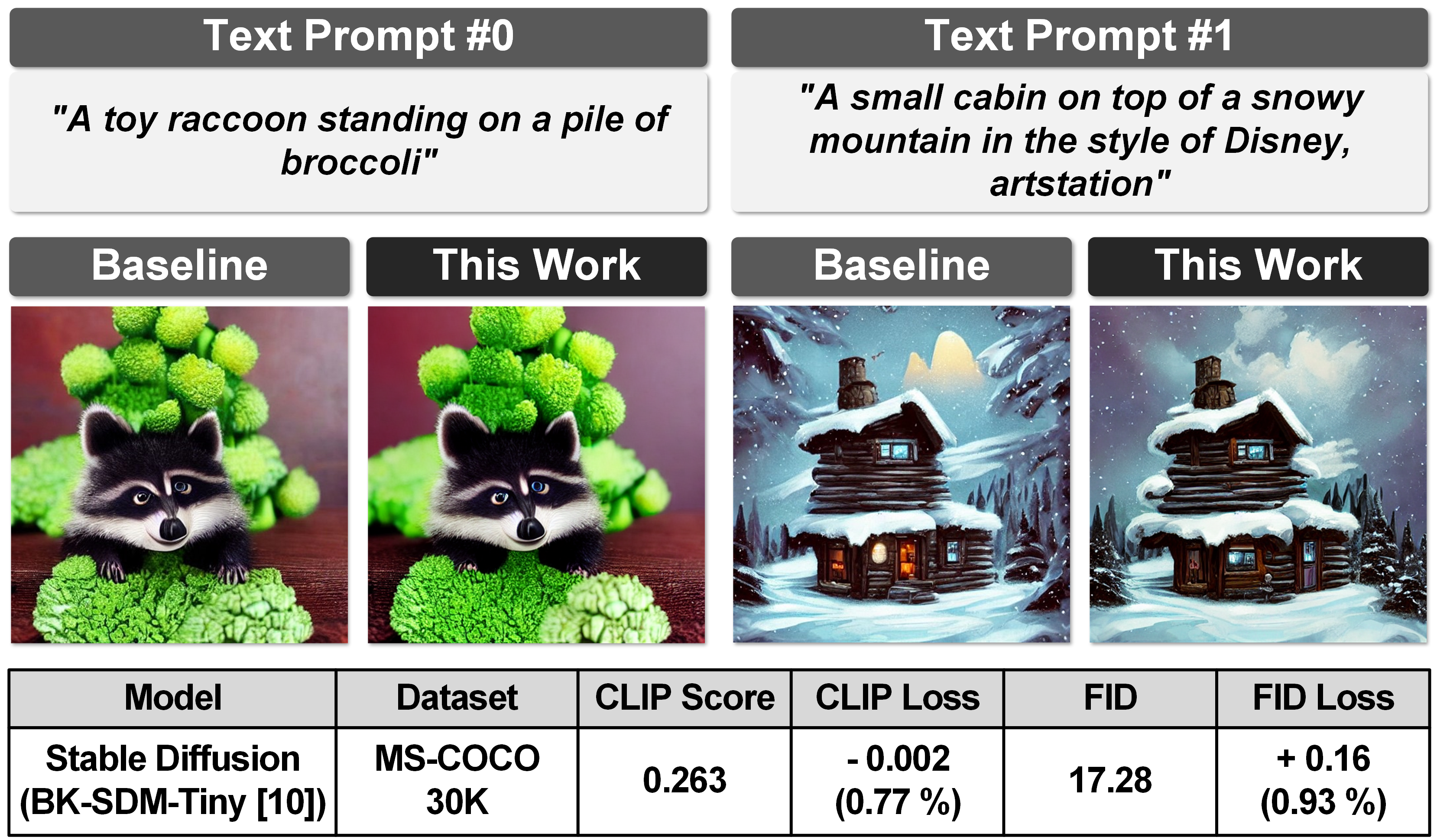}}
\caption{Text-to-image generation results and evaluation.}
\label{fig11}
\vspace{-2mm}
\end{figure}

Fig.~\ref{fig10} shows the layout photograph of the proposed SD processor, which is implemented in 28 nm CMOS technology and occupies a die area of 3.15 \(mm^2\). It consumes an average power of 225.6 mW and a peak throughput of 3.84 TOPS under 1 V supply voltage and 250 MHz clock frequency. As a result, the proposed SD processor achieved an energy efficiency of 28.6 and 213.3 mJ/iteration, respectively, for cases without and with consideration of EMA. Fig.~\ref{fig11} displays the text-to-image generation results, including generated images and benchmark evaluation. When evaluated on BK-SDM \cite{b10} network architecture and MS-COCO dataset, the processor successfully generated images from the given texts with only \textless 1 \% CLIP Loss and FID Loss. Table.~\ref{table1} shows the comparisons with previous transformers and generative AI processors. Being the sole processor optimized for SD, it not only achieved a higher peak energy efficiency of 14.94 TOPS/W compared to previous works \cite{b12, b14} but also reduced the EMA-included energy consumption per iteration by 34.6 \% with the proposed EMA and computation reduction schemes.

\begin{table}[tp]
\caption{Comparison Table}
\centerline{\includegraphics[width=\columnwidth]{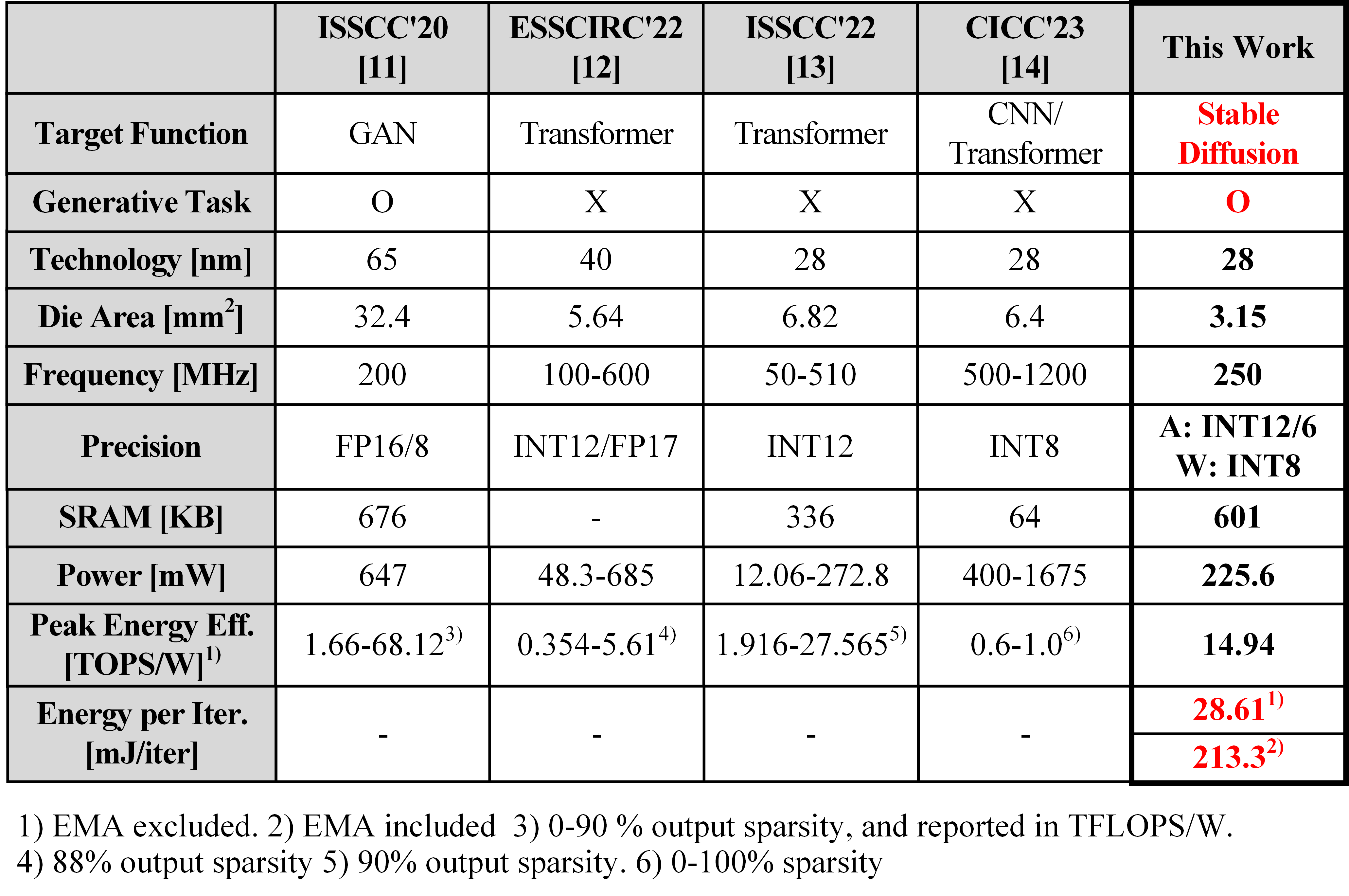}}
\label{table1}
\vspace{-7mm}
\end{table}

\section{Conclusion}
An energy-efficient stable diffusion processor for text-to-image generation is proposed. It enhances the throughput and energy efficiency by following key features: 1) Patch similarity-based sparsity augmentation (PSSA) reduces the EMA energy of SAS by 60.3 \% and its index overhead by 83.6 \%. 2) Text-based important pixel spotting (TIPS) allows 44.8 \% of the FFN layer workload to be processed with low-precision activation. 3) Dual-mode bit-slice core (DBSC) architecture for supporting mixed-precision computation and layer-aware dual-stationary-mode increases 43.0 \% of energy efficiency in FFN layers. Overall, 28.6 mJ/iteration of high energy efficiency could be achieved on the MS-COCO dataset with high-quality image generation of only 0.002 CLIP score loss and 0.16 FID loss.

\section*{Acknowledgment}
The EDA tool was supported by the IC Design Education Center (IDEC), Korea.

\vspace{12pt}
\end{document}